\documentclass{desyproc}

\usepackage[utf8]{inputenc}
\usepackage{subcaption}
\usepackage{multirow}
\begin{document}
\title{Light Collection in the Prototypes of the ANAIS Dark Matter Project}

\author{{\slshape  J. Amaré$^{ 1,2}$ , S. Cebrián$^{ 1,2}$ , C. Cuesta$^{1,2}$\footnote{Present address: CENPA and Department of Physics, University of Washington, Seattle, WA, USA} , E. García$^{ 1,2}$ , M. Martínez$^{ 1,2}\footnote{Present address: Univ. Roma La Sapienza, Roma, Italy}$ , M.A. Oliván$^{ 1,2}\footnote{Corresponding author (e-mail: maolivan@unizar.es)}$, Y. Ortigoza$^{ 1,2}$ , A. Ortiz de Solórzano$^{ 1,2}$ , C. Pobes$^{1,2}$\footnote{Present address: Instituto de Ciencia de Materiales de Aragón, CSIC - Universidad de Zaragoza, Spain} , J. Puimedón$^{ 1,2}$ , M.L. Sarsa$^{ 1,2}$ , \mbox{J.A. Villar$^{ 1,2}$ ,} P. Villar$^{ 1,2}$}\\[1ex]
$^1$Laboratorio de Física Nuclear y Astropartículas, Universidad de Zaragoza, Pedro Cerbuna 12, 50009, Zaragoza, Spain,\\
$^2$Laboratorio Subterráneo de Canfranc, Paseo de los Ayerbe s/n, 22880 Canfranc Estación, Huesca, Spain\\
}

\contribID{familyname\_firstname}

\confID{11832}  
\desyproc{DESY-PROC-2015-02}
\acronym{Patras 2015} 
\doi  

\maketitle

\begin{abstract}The ANAIS experiment aims at the confirmation of the DAMA/LIBRA signal using the same target and technique at the Canfranc Underground Laboratory (LSC) in Spain. ANAIS detectors consist of large NaI crystals coupled to two photomultipliers (PMTs). In this work we present Single Electron Response (SER) data for several units of the Hamamatsu R12669SEL2 PMT model extracted from normal operation data of ANAIS underground prototypes and we compare them with PMT SER characterization previously done at surface lab before coupling them to NaI crystal. Moreover, total light collection for different ANAIS prototypes has been calculated, producing an excellent average result of 15 phe/keV, which has a good impact in both energy resolution and threshold.
\end{abstract}

\section{Introduction}\label{sec:LY}
The ANAIS (Annual Modulation with NaI(Tl) Scintillators) experiment~\cite{ANAIS-25, TWA-250} is intended to confirm the DAMA/LIBRA signal~\cite{DAMA-fin} using the same target and technique at the Canfranc Underground Laboratory. The ANAIS-25 set-up consisted of two NaI(Tl) detectors of 12.5 kg each manufactured by Alpha Spectra (named D0 and D1 in this work). It has been taking data since December 2012 in order to measure the internal contamination of the NaI(Tl) crystals and assess the performance of the detectors. A new Alpha Spectra detector (named D2 in this work) with lower internal background~\cite{ANABKG} was received in March 2015, which added to ANAIS-25 modules formed the ANAIS-37 set-up. Every detector has been coupled to two Hamamatsu R12669SEL2 PMTs, the ANAIS selected model~\cite{CCUESTA}. In the following we will report on the PMT Single Electron Response (SER) data extracted from both set-ups on underground site and along normal operation which have been compared with the SER characterization of the PMTs previously performed at Zaragoza (Section 2) and on the estimates of the total light collection for all the available detectors (Sections 3 and 4).
\section{SER extraction}
\begin{figure}[h]
     \begin{center}
        \begin{subfigure}[b]{0.32\textwidth}
        \includegraphics[width=\textwidth]{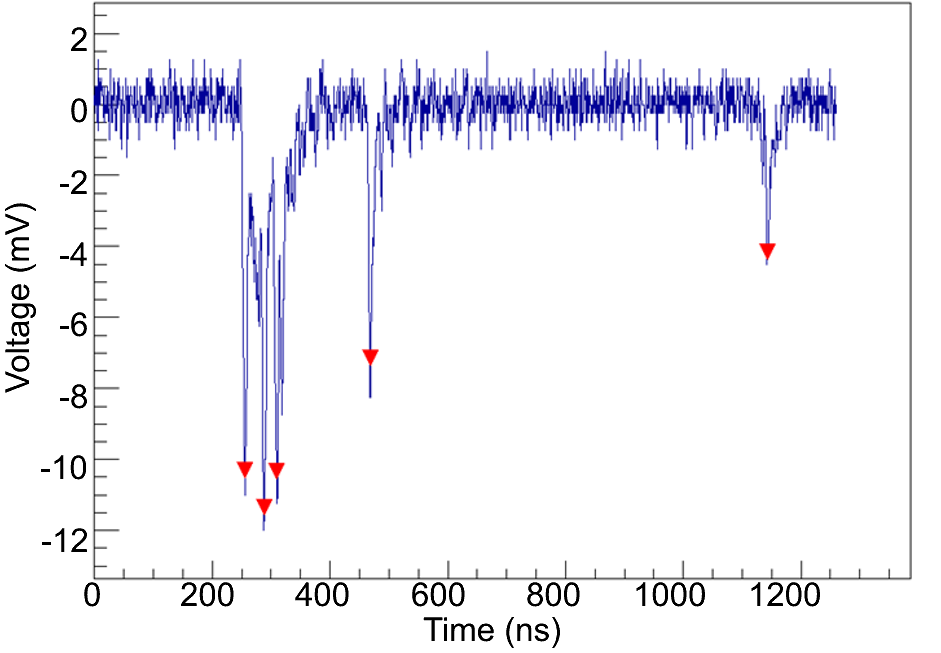}
	\caption{\label{fig:LP}}
        \end{subfigure}%
	~
	\begin{subfigure}[b]{0.33\textwidth}
        \includegraphics[width=.95\textwidth]{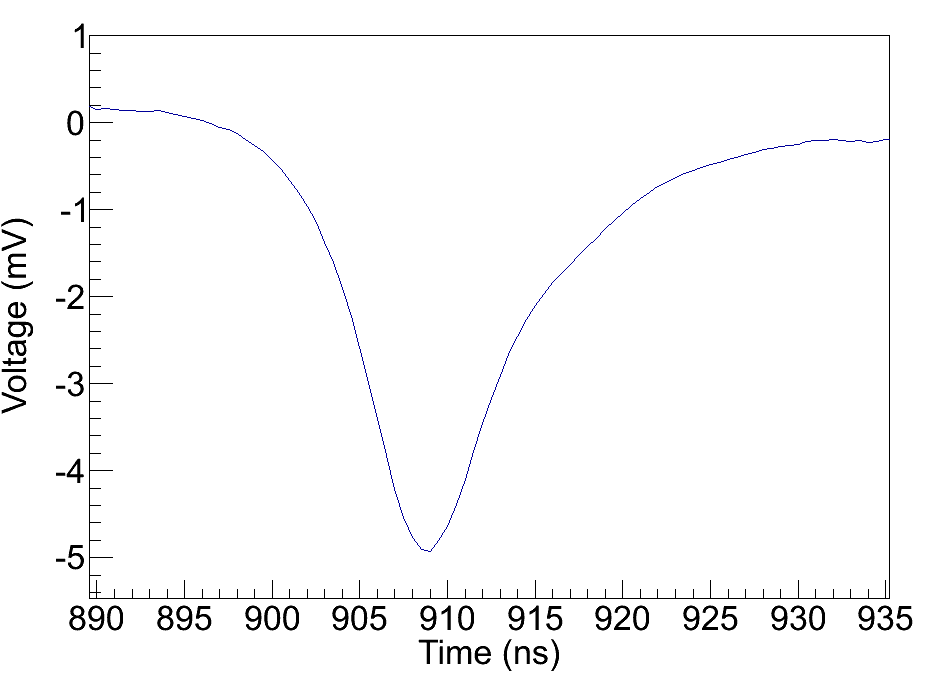}
	\caption{\label{fig:LPMean}}
        \end{subfigure}%
	~
	\begin{subfigure}[b]{0.33\textwidth}
        \includegraphics[width=.95\textwidth]{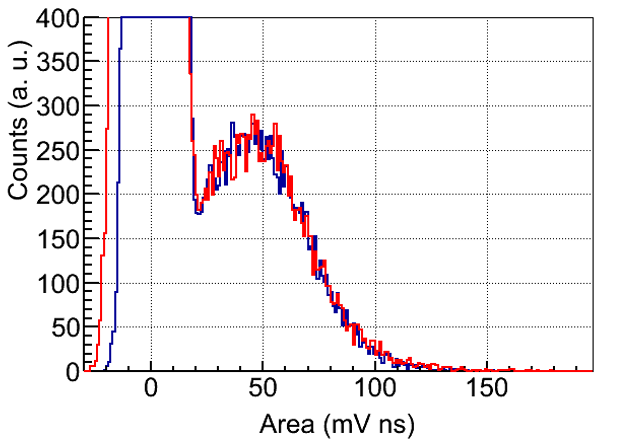}
	\caption{\label{fig:SER_LP_LED}}
        \end{subfigure}%

	\caption[]{Pulse with a low number of phe; peaks identified by the applied algorithm are shown with red triangles (a), SER mean pulse (b) and SER charge distributions derived at PMT test bench (red) and along normal operation (blue) (c)\label{fig:LPPlots}.}
\end{center}
\end{figure}

First, the PMTs SER was measured at the Zaragoza test bench using UV LED illumination of very low intensity, and triggering in the excitation LED signal. This characterization was done for each PMT unit before mounting ANAIS detectors, and allowed to validate the SER determination onsite along normal operation of ANAIS detectors at the LSC. The latter was performed applying a peak identification algorithm to the data taken at the LSC, building the SER by selecting peaks at the end of the pulse of each PMT in order to avoid trigger bias in pulses having a low number of peaks to prevent the pile-up of several photoelectrons (phe). An example of a pulse fulfilling these conditions can be seen in Figure \ref{fig:LP} and the mean pulse of a selection of this kind of events is shown in Figure \ref{fig:LPMean}. The phe area (proportional to charge) is integrated in a fixed time window around the peak maximum in order to obtain the single electron response charge distribution. The SER charge distribution extracted for the same PMT by these two methods is compared in Figure \ref{fig:SER_LP_LED} showing full agreement between both.
\section{ANAIS-25}
The light collected by each of the PMTs coupled to the ANAIS-25 modules was calculated by dividing the mean value of the charge distribution associated to a known energy deposition in the NaI crystal and the mean value of the SER charge distribution derived as aforementioned. The 22.6 keV line from a $^{109}Cd$ calibration source was used for this study. The result of the SER charge spectrum and the $^{109}Cd$ line Gaussian fits can be seen in Table~\ref{tab:A25_Fits} (PMT ij corresponds to PMT j of detector Di). These results and the global light collection of the two ANAIS-25 detectors are summarized in Table \ref{tab:A25LY}. The $^{109}Cd$ line resolution is also calculated and is shown in Table \ref{tab:A25CdRes}. These results confirm the prototypes outstanding light collection and its impact in resolution. The very good optical performance of the Alpha Spectra modules evidenced by these figures is very promising in order to reduce the energy threshold below 2 keVee~\cite{ANAISPATRAS}. 
\begin{table}[h!]
		\begin{center}
			\begin{tabular}{| c  c  c  c  c|}
				\hline
				PMT& \begin{tabular}{@{}c@{}} SER mean\\mV·ns \end{tabular}& \begin{tabular}{@{}c@{}}SER $\sigma$\\mV·ns\end{tabular} & \begin{tabular}{@{}c@{}}22.6 keV mean\\mV·ns \end{tabular}& \begin{tabular}{@{}c@{}}22.6 keV\ $\sigma$\\mV·ns \end{tabular}\\
				\hline
				 00&$35.47\pm0.35$ & $21.73\pm0.25$ & $6122\pm2$ & $669\pm2$\\
				 01&$29.42\pm0.21$ & $18.81\pm0.24$ & $5057\pm2$ & $568\pm2$\\
				10&$41.20\pm0.25$ & $28.30\pm0.21$ & $7139\pm4$ & $809\pm4$\\
				11&$44.52\pm0.29$ & $24.36\pm0.24$ & $7570\pm4$ & $825\pm3$\\
				\hline

			\end{tabular}
			\caption[ANAIS-25 values for SER charge distribution and $^{109}Cd$ 22.6 keV line Gaussian fits]{ANAIS-25 values for SER charge distribution and $^{109}Cd$ 22.6 keV line Gaussian fits.} 
			\label{tab:A25_Fits} 
		\end{center}
	\end{table}
	\begin{table}[h!]
		\parbox{.45\linewidth}{
		\begin{center}
			\begin{tabular}{|ccc|}
				\hline
				PMT& \begin{tabular}{@{}c@{}}PMT\\ phe/keV\end{tabular}& \begin{tabular}{@{}c@{}}Detector\\ phe/keV\end{tabular}\\
				\hline
				00& $7.64\pm0.08$&\multirow{2}{*}{$15.24\pm0.09$}\\
				01&$7.61\pm0.05$& \\
				10& $7.67\pm0.05$&\multirow{2}{*}{$15.19\pm0.07$} \\
				11&$7.52\pm0.05$ & \\
				\hline

			\end{tabular}
			\caption[ANAIS-25 light collection]{ANAIS-25 light collection.} 
			\label{tab:A25LY} 
		\end{center}
	}
	\hspace{.9em}
	\parbox{.45\linewidth}{
		\begin{center}
			\begin{tabular}{|ccc|}
				\hline
				PMT& \begin{tabular}{@{}c@{}}PMT \\ $\sigma$/E (\%)\end{tabular}&\begin{tabular}{@{}c@{}}Detector\\$\sigma$/E (\%)\end{tabular}\\
				\hline
				00& $10.93\pm0.03$&\multirow{2}{*}{$8.51\pm0.03$}\\
				01&$11.24 \pm 0.04$& \\
				10& $11.33\pm0.05$ &\multirow{2}{*}{$8.59\pm0.04$}\\
				11&$10.90\pm0.05$&\multicolumn{-2}{c|}{} \\
				\hline

			\end{tabular}
			\caption[ANAIS-25 resolution at 22.6 keV]{ANAIS-25 resolution at 22.6 keV.} 
			\label{tab:A25CdRes} 
		\end{center}
	}

	\end{table}
\section{ANAIS-37}
The same procedure was repeated with ANAIS-37 setup data. In this setup the operating voltages of the D0 and D1 detectors were increased in order to better study the low energy region and for this reason the SER values are higher. The voltage of the new detector (D2) was selected to have a $10^6$ gain value in both PMTs in order to explore the high energy region~\cite{ANAISPATRAS}. The results of the SER charge distribution and the $^{109}Cd$ 22.6 keV line Gaussian fits can be seen in Table \ref{tab:A37_Fits}. The light collection for every PMT and detector can be observed in Table \ref{tab:A37LY}. The newly extracted values for D0 and D1 are compatible with those obtained for ANAIS-25 (see previous section). Good values for the new D2 ($\sim 16$ phe/keV) have also been measured having again a good impact in terms of energy threshold and resolution, crucial for the sensitivity to WIMPs annual modulation. 
\begin{table}[h!]
		\begin{center}
			\begin{tabular}{| c  c  c  c  c |}
				\hline
				PMT& \begin{tabular}{@{}c@{}} SER mean\\mV·ns \end{tabular}& \begin{tabular}{@{}c@{}}SER $\sigma$\\mV·ns\end{tabular} & \begin{tabular}{@{}c@{}}22.6 keV mean\\mV·ns \end{tabular}& \begin{tabular}{@{}c@{}}22.6 keV $\sigma$\\mV·ns \end{tabular}\\
				\hline
				 00&$61.47\pm0.36$ & $35.02\pm0.32$ & $10257\pm5$ & $1126\pm4$\\
				 01&$58.40\pm0.71$ & $43.06\pm0.51$ & $10425\pm5$ & $1166\pm4$\\
				10&$83.24\pm0.55$ & $46.52\pm0.51$ & $12820\pm5$ & $1463\pm4$\\
				11&$73.91\pm0.74$ & $42.04\pm0.52$ & $12740\pm5$ & $1404\pm4$\\
				20&$42.70\pm2.10$ & $25.42\pm1.79$ & $7928\pm5$ & $909\pm6$\\
				21&$44.57\pm2.10$ & $26.67\pm1.95$ & $8155\pm6$ & $930\pm6$\\
				\hline
			\end{tabular}
			\caption[ANAIS-37 value from SER charge distribution and $^{109}Cd$ 22.6 keV line Gaussian fits]{ANAIS-37 values from SER charge distribution and $^{109}Cd$ 22.6 keV line Gaussian fits.} 
			\label{tab:A37_Fits} 
		\end{center}
	\end{table}
	\begin{table}[h!]
		\parbox{.45\linewidth}{
		\begin{center}
			\begin{tabular}{|ccc|}
				\hline
				PMT& \begin{tabular}{@{}c@{}}PMT\\ phe/keV\end{tabular}& \begin{tabular}{@{}c@{}}Detector\\ phe/keV\end{tabular}\\
				\hline
				00& $7.38\pm0.04$&\multirow{2}{*}{$15.26\pm0.10$}\\
				01&$7.88\pm0.09$& \\
				10& $6.81\pm0.05$&\multirow{2}{*}{$14.44\pm0.09$} \\
				11&$7.62\pm0.08$ & \\
				20& $8.21\pm0.40$&\multirow{2}{*}{$16.31\pm0.56$} \\
				21&$8.09\pm0.38$ & \\
				\hline

			\end{tabular}
			\caption[ANAIS-37 light collection]{ANAIS-37 light collection.} 
			\label{tab:A37LY} 
		\end{center}
	}
	\hspace{.9em}
	\parbox{.45\linewidth}{
		\begin{center}
			\begin{tabular}{|ccc|}
				\hline
				PMT& \begin{tabular}{@{}c@{}}PMT\\ $\sigma$/E (\%)\end{tabular}&\begin{tabular}{@{}c@{}}Detector\\$\sigma$/E (\%)\end{tabular}\\
				\hline
				00& $10.97\pm0.04$&\multirow{2}{*}{$8.73\pm0.03$}\\
				01&$11.18 \pm 0.04$& \\
				10& $11.40\pm0.03$ &\multirow{2}{*}{$8.80\pm0.03$}\\
				11&$11.02\pm0.03$& \multicolumn{-2}{c|}{}\\
				20& $11.46\pm0.07$ &\multirow{2}{*}{$8.99\pm0.05$}\\
				21&$11.40\pm0.08$ &\multicolumn{-2}{c|}{}\\
				\hline

			\end{tabular}
			\caption[ANAIS-37 resolution at 22.6 keV]{ANAIS-37 resolution at 22.6 keV.} 
			\label{tab:A37CdRes} 
		\end{center}
	}

	\end{table}
	\section{Conclusion}
	The PMTs single electron response was characterized along detectors normal operation and compared with the previous PMTs measurements showing a full compatibility among them. Using this extraction, an excellent light collection for the three ANAIS detectors, of the order of $\sim$15 phe/keV, has been measured. Thanks to this, an energy threshold for the ANAIS experiment at 1 keVee is at reach, depending now on improving the filtering protocols for PMT origin coincident events, which would significantly improve the sensitivity of the ANAIS Project in the search for the annual modulation effect in the WIMPs signal~\cite{ANAISPATRAS}.
	\section{Acknowledgments}
This work was supported by the Spanish Ministerio de Economía y Competitividad and the European Regional Development Fund (MINECO-FEDER) (FPA2014-55986), the Consolider-Ingenio 2010 Programme under grants MULTIDARK CSD2009-00064 and CPAN CSD2007-00042, and the Gobierno de Aragón (GIFNA and ARAID Foundation). P. Villar is supported by the MINECO Subprograma de Formación de Personal Investigador. We also acknowledge LSC and GIFNA staff for their support.

\begin{footnotesize}

\end{footnotesize}


\end{document}